\documentclass[%
 aip,amsmath,amssymb,
 reprint]{revtex4-1}

\usepackage{graphicx}
\usepackage{dcolumn}
\usepackage{bm}

\usepackage{chemformula}
\usepackage{gensymb}
\usepackage{hyperref}

\usepackage[utf8]{inputenc}
\usepackage[T1]{fontenc}
\usepackage{mathptmx}
\usepackage{etoolbox}

\makeatletter
\def\@email#1#2{%
 \endgroup
 \patchcmd{\titleblock@produce}
  {\frontmatter@RRAPformat}
  {\frontmatter@RRAPformat{\produce@RRAP{*#1\href{mailto:#2}{#2}}}\frontmatter@RRAPformat}
  {}{}
}%
\makeatother
\begin{document}

\title[Microstructural and preliminary optical and microwave characterization of erbium doped \ch{CaMoO4} thin films]{Microstructural and preliminary optical and microwave characterization of erbium doped \ch{CaMoO4} thin films}

\author{Ignas Masiulionis}
\affiliation{Pritzker School of Molecular Engineering, University of Chicago, Chicago, Illinois 60637, United States.}
\affiliation{Materials Science Division, Argonne National Laboratory, Lemont, Illinois 60439, United States.}
\author{Bonnie Y.X. Lin}
\affiliation{Department of Materials Science and Engineering, Massachusetts Institute of Technology, Cambridge, Massachusetts 02139, United States.}
\author{Sagar Kumar Seth}
\affiliation{Pritzker School of Molecular Engineering, University of Chicago, Chicago, Illinois 60637, United States.}
\affiliation{Materials Science Division, Argonne National Laboratory, Lemont, Illinois 60439, United States.}
\author{Gregory D. Grant}
\affiliation{Pritzker School of Molecular Engineering, University of Chicago, Chicago, Illinois 60637, United States.}
\affiliation{Materials Science Division, Argonne National Laboratory, Lemont, Illinois 60439, United States.}
\author{Wanda L. Lindquist}
\affiliation{Materials Science Division, Argonne National Laboratory, Lemont, Illinois 60439, United States.}
\affiliation{Open Quantum Initiative, Chicago Quantum Exchange, Chicago, Illinois 60615, United States.}
\author{Sungjoon Kim}
\affiliation{Applied Materials Division, Argonne National Laboratory, Lemont, Illinois 60439, United States.}
\author{Junghwa Kim}
\affiliation{Department of Materials Science \& Engineering, Massachusetts Institute of Technology, Cambridge, Massachusetts 02139, United States.}
\author{Angel Yanguas-Gil}
\affiliation{Applied Materials Division, Argonne National Laboratory, Lemont, Illinois 60439, United States.}
\author{Jeffrey W. Elam}
\affiliation{Applied Materials Division, Argonne National Laboratory, Lemont, Illinois 60439, United States.}
\author{Jiefei Zhang}
\affiliation{Materials Science Division, Argonne National Laboratory, Lemont, Illinois 60439, United States.}
\affiliation{Center for Molecular Engineering, Argonne National Laboratory, Lemont, IL, 60439, United States.}
\author{James M. LeBeau}
\affiliation{Department of Materials Science \& Engineering, Massachusetts Institute of Technology, Cambridge, Massachusetts 02139, United States.}
\author{David D. Awschalom}
\affiliation{Pritzker School of Molecular Engineering, University of Chicago, Chicago, Illinois 60637, United States.}
\affiliation{Materials Science Division, Argonne National Laboratory, Lemont, Illinois 60439, United States.}
\affiliation{Department of Physics, University of Chicago, Chicago, Illinois 60637, United States.}
\affiliation{Center for Molecular Engineering, Argonne National Laboratory, Lemont, IL, 60439, United States.}
\author{Supratik Guha}
 \email{guha@uchicago.edu.}
\affiliation{Pritzker School of Molecular Engineering, University of Chicago, Chicago, Illinois 60637, United States.}
\affiliation{Materials Science Division, Argonne National Laboratory, Lemont, Illinois 60439, United States.}
\affiliation{Center for Molecular Engineering, Argonne National Laboratory, Lemont, IL, 60439, United States.}

\date{\today}
             
\hrulefill

\begin{abstract}
This work explores erbium-doped calcium molybdate (\ch{Er:CaMoO4}) thin films grown on silicon and yttria stabilized zirconia (YSZ) substrates, as a potential solid state system for C-band (utilizing the $\sim$1.5 $\mu$m Er$^{3+}$ 4f-4f transition) quantum emitters for quantum network applications. Through molecular beam epitaxial growth experiments and electron microscopy, X-ray diffraction and reflection electron diffraction studies, we identify an incorporation limited deposition regime that enables a 1:1 Ca:Mo ratio in the growing film leading to single phase \ch{CaMoO4} formation that can be in-situ doped with Er (typically 2-100 ppm). We further show that growth on silicon substrates is single phase but polycrystalline in morphology; while growth on YSZ substrates leads to high-quality epitaxial single crystalline \ch{CaMoO4} films. We perform preliminary optical and microwave characterization on the suspected $Y_1 - Z_1$ transition of 2 ppm, 200 nm epitaxial \ch{Er:CaMoO4} annealed thin films and extract an optical inhomogeneous linewidth of 9.1(1) GHz, an optical excited state lifetime of 6.7(2) ms, a spectral diffusion-limited homogeneous linewidth of 6.7(4) MHz, and an EPR linewidth of 1.10(2) GHz. 
\end{abstract}

\hrulefill

\maketitle

\section{\label{sec:level1}Introduction}

Erbium doped solid-state materials are potential candidates as memories for quantum repeaters due to its $^4I_{13/2} - ^4I_{15/2}$ optical transition at 1.5 $\mu$m lying within the telecom C-band\cite{wolfowicz2021}. Rare-earth oxides have been explored as promising hosts for erbium \cite{dibos2022, ji2024, singh2020, horvath2023, rinner2023, zhang2024, grant2024} with long spin coherence times. The intrinsic spin-photon interface and long coherence times of erbium in these hosts strengthen them as viable platforms.

Out of the various rare-earth oxide platforms, Er doped \ch{CaWO4} has emerged as one of the most promising candidates. Spin coherence times of 23 ms at 10 mK for \ch{Er^{3+}} spin transition, limited by nuclear spin bath~\cite{le2021, kanai2022}, and indistinguishable single-photon generation~\cite{ourari2023} have been demonstrated in bulk \ch{Er:CaWO4} crystals. The closely related compound \ch{CaMoO4}, which possesses the same scheelite crystal structure, has the potential to host \ch{Er^{3+}} ions with comparable coherence properties given its similar natural abundance spin active isotope concentration ($^{183}$W of 14.3\% vs. $^{95}$Mo and $^{97}$Mo of 25.5\%). The magnetic and optical properties of \ch{Er^{3+}} ions in bulk \ch{CaMoO4} have been reported in a recent study~\cite{gerasivom2024}.

    The above studies on \ch{Er^{3+}} in bulk crystals of \ch{CaWO4} and \ch{CaMoO4} have established the feasibility of these materials for quantum network applications. Our interests are to explore the epitaxially grown thin film form of these compounds for on-chip integration and scalable photonic quantum memory devices. \ch{CaMoO4} and \ch{CaWO4} possess a scheelite structure with the $C^6_{4h}$ space symmetry and lattice constants of $a=b=5.2236$ \r{A} and $c=11.4285$ \r{A}~\cite{murai2023}, and $a=b=5.24$ \r{A} and $c=11.37$ \r{A}~\cite{le2021} respectively. Therefore, both of these compounds are expected to possess epitaxial compatibility with substrates such as silicon (lattice mismatch of -3.8\%) and yttria stabilized zirconia (YSZ) (lattice mismatch of +1.9\%). On the other hand, maintaining a consistent 1:1 Ca:Mo ratio in the deposited thin film will ideally require the development development of a compositionally self limiting growth process controlled by surface incorporation.

We have undertaken a systematic exploration of the molecular beam epitaxial (MBE) growth and properties of epitaxial thin films of these compounds, with a focus on \ch{CaMoO4} thin films, the topic of the work reported here.  Through a detailed study of the MBE growth process combined with X-ray diffraction and scanning transmission electron microscopy (STEM) studies, we have identified a surface incorporation limited deposition regime that yields single phase epitaxial films of \ch{CaMoO4} on YSZ substrates. We show that when we grow on Si(001) substrates, epitaxial registry is lost, leading to polycrystalline films. Detailed optical and magnetic characterization on the 2 ppm \ch{Er:CaMoO4} annealed at 1000\celsius{} in 20\% \ch{Ar/O2} ambient for one hour have been performed, with respect to the suspected Er$^{3+}$ $Y_1-Z_1$ transition. We report optical inhomogeneous linewidths ($\Gamma_{inh}$) of 9.1(1) GHz, optical relaxation time ($T_1$) of 6.7(2) ms, transient spectral holeburning linewidth ($\Gamma_{SHB}$) of 6.7(4) MHz, magnetic paramagnetic resonance linewidth ($\Gamma_{EPR}$) of 1.10(2) GHz, and g-factor of 8.261(3), results that are comparable with other Er doped epitaxially grown thin film oxides such as \ch{CeO2}~\cite{grant2024}.

\begin{figure*}[t]
    \centering
    \includegraphics{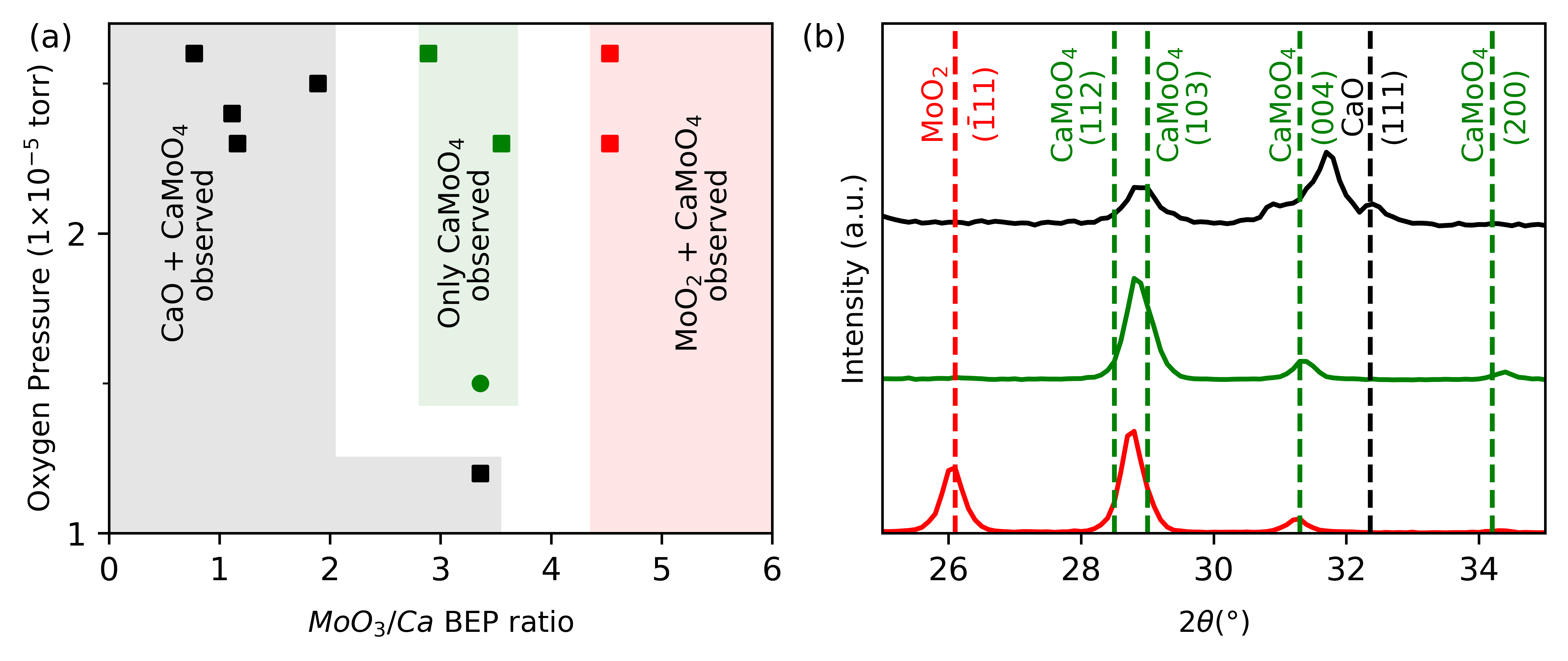}
    \caption{Stoichiometric growth window determination of \ch{CaMoO4} on silicon (001). (a) Growth window dependence on oxygen pressure and \ch{MoO3}/\ch{Ca} BEP ratio. Area in light black indicate growth window yielding \ch{CaMoO4} and \ch{CaO} phases, area in red indicate growth window with \ch{CaMoO4} and \ch{MoO2}, and lastly green area shows the desired growth window of stoichiometric \ch{CaMoO4}. Round markers indicate growths performed using molecular \ch{O2}, while squares show growths performed with atomic oxygen. The growths demonstrated here are performed at substrate temperatures 650\celsius{}. (b) GIXRD comparison between growths in three different regions described in (a). For growths in red area, there is a prominent \ch{MoO2} ($\bar{1}$11) peak at 26.033$\degree$, growths in black area, prominent \ch{CaO} (111) peak at 32.204$\degree$, and the growths in green region did not yield any of the \ch{MoO2} and \ch{CaO} diffraction peaks.}
    \label{SGW}
   
\end{figure*}

\section{Methods}

\subsection{\ch{Er:CaMoO4} thin film growth and substrate preparation}

In order to grow \ch{CaMoO4} via MBE, we use metallic \ch{Ca} (chunks, 99.99\% trace metal basis), powdered \ch{MoO3} (ThermoScientific Molybdenum(VI) oxide, Puratronic\textsuperscript \textregistered, 99.9995\% metal basis excluding W), and ultra-high purity oxygen (Matheson, 99.9999\%). Solid sources are loaded into RIBER ABN60DF dual-zone low temperature effusion cells, mounted on a RIBER CZ21 MBE Cluster reactor. 

Typical range of operational temperatures for calcium are 380-520 \celsius{} and molybdenum trioxide are 560-620 \celsius{}. A molecular or atomic oxygen flux is delivered through a MKS mass flow controller at flow rates between 0.5-3.0 sccm (leading to a beam equivalent pressure of 5e-6 to 3e-5 torr). Some growths use atomic oxygen that was generated using a RIBER RF oxygen plasma source, with power range of $325-425W$. 

During deposition, substrate temperatures are kept in the range of 650-675 \celsius{} as measured by a pyrometer. Films are doped with 2-100 ppm erbium using a HT-12 RIBER effusion cell~\cite{grant2024, singh2020} for optical and magnetic characterization. An ion gauge in line of sight to the effusion cells is used to measure fluxes (referred to as beam equivalent pressure, BEP) and the growth surface is monitored \textit{in-situ} using a reflection high-energy electron diffraction (RHEED) system operated at 15 kV. 

Silicon (001) substrates are prepared using a modified HF-last Radio Corporation of America (RCA)~\cite{reinhardt2018} cleaning process, outgassed at 300 \celsius{} for 30 minutes in a dedicated ultrahigh-vacuum outgassing chamber, followed by in-vacuum transfer into the deposition chamber. The silicon substrate is then heated to 750 \celsius{} and annealed for 5 minutes to achieve Si (001) 2 x 1 reconstruction, following which it is brought to the growth temperature. Prior to the growth, the oxygen flow is turned on and the oxygen plasma is triggered (in the cases where atomic oxygen is used). 

Growth is initiated by simultaneously opening \ch{Ca}, \ch{MoO3}, and oxygen shutters and after 5 seconds, opening the substrate shutter. Following $\sim10$ nm of undoped \ch{CaMoO4} buffer layer growth, the erbium shutter is opened for the deposition of \ch{Er:CaMoO4} thin films  with thicknesses of a few hundreds of nanometers. Following the doped layer growth, an additional $\sim 10$nm of undoped \ch{CaMoO4} capping layer is deposited~\cite{singh2024}. Following the deposition, the films are cooled in the presence of molecular oxygen. 

Post growth anneals are performed in a MTI OTF-1200X tube furnace in 1 atm of \ch{O2/Ar} (20/80 mix) with 50 mL/min flow rate for 1 hr set at 1000 \celsius{} temperatures ranging from 550 to 1000 \celsius{}.

For the growth on YSZ, the YSZ substrates are prepared by 10-minute acetone followed by 10-minute isopropanol sonication bath at room temperature. The substrates are then loaded into Riber C21 DZ Cluster MBE and degassed for 30 minutes at 300 \celsius{},  prior to introduction into the reactor chamber. The surface is further cleaned for 20 minutes under 1.5e-5 torr of atomic oxygen with $325W$ prior to heating it to growth temperature, resulting in a smooth and atomically sharp surface.

\subsection{Structural and optical characterization}

\begin{figure*}[t]
    \centering
    \includegraphics{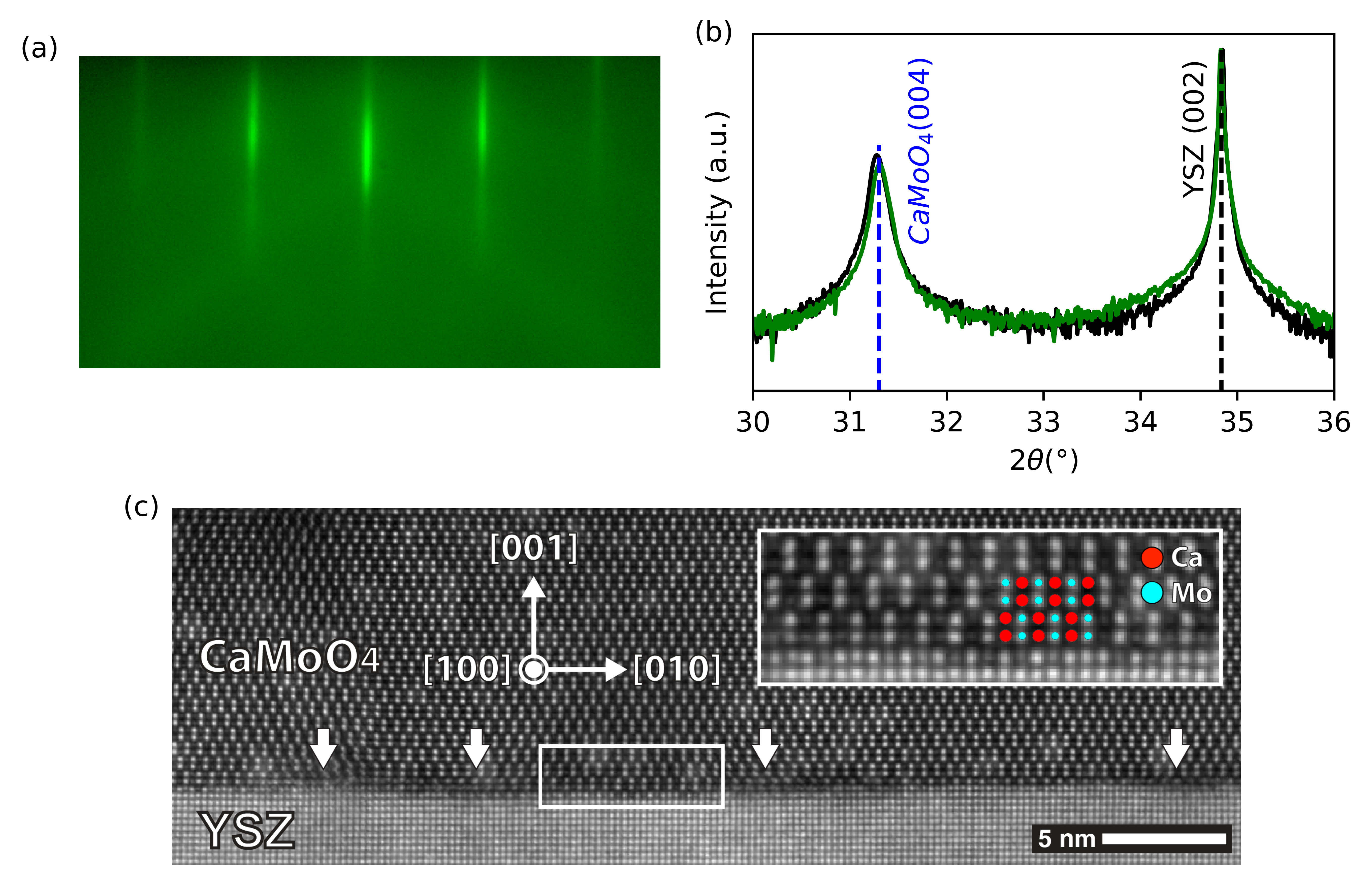}
    \caption{Epitaxy of erbium doped 200 nm \ch{CaMoO4} on a YSZ (001) substrate. (a) RHEED pattern post growth along YSZ <100> azimuthal direction, showing sharp streaks, indicating high quality epitaxy. (b) $\omega$-2$\Theta$ HRXRD scans of as-grown (in black) and annealed (in green) \ch{Er:CaMoO4}, showing the \ch{CaMoO4} (004) diffraction peak located at 31.292$\degree$. (c) Cross-sectional atomic resolution HAADF-STEM image of the atomically sharp annealed \ch{CaMoO4}/YSZ interface, viewed along  [100]. Inset shows a magnified view of the interface overlapped with a schematic scheelite structure (only cation positions are shown). Misfit dislocations are marked by white arrows. }
    \label{EG}

\end{figure*}

A two-circle Bruker D8 Discover diffractometer operated with micro-focus copper X-ray source (I$\mu$S) at 50 kV and 1000 $\mu$A is used to perform grazing incident x-ray diffraction (GIXRD) scans aligned to the \ch{CaMoO4} (112) reflection located at 28.701\degree{} 2$\theta$, and high-resolution x-ray diffraction (HRXRD) scans, which are aligned to YSZ (002) reflection located at 34.835\degree{} 2$\theta$.

SEM was carried out using a JEOL IT800 instrument with an acceleration of 10 kV.

Continuous wave (CW) X-band (9.385 GHz) electron paramagnetic resonance (EPR) experiments are carried out using a Bruker ELEXSYS II E500 EPR spectrometer (Bruker BioSpin), equipped with a TE102 rectangular EPR resonator (Bruker ER 4102ST). Field modulation at 100 kHz in combination with lock-in detection leads to first derivative-type CW EPR spectra. The measurements are performed at cryogenic temperatures between 4.3 and 4.6 K, with temperature governed by a helium gas-flow cryostat (ICE Oxford) and an ITC (Oxford Instruments). The \ch{Er:CaMoO4} samples are mounted with the static magnetic field parallel to the YSZ⟨100⟩ axis. The data are collected using a field modulation of 1 mT, a microwave power attenuation of 25 dB (from 200 mW).

Cross-sectional samples for electron microscopy are prepared along \ch{YSZ}[100] using a FEI Helios 600i DualBeam scanning electron microscopy (SEM)/focused ion beam (FIB), followed by Ar$^+$ ion milling with a Fischione 1051 TEM Mill. An aberration-corrected Thermo Fisher Scientific Themis Z scanning/transmission electron microscope is operated at 200 kV for STEM imaging, energy dispersive spectroscopy (EDS), and four-dimensional (4D) nano-beam electron diffraction (NBED). For STEM imaging, the probe has a convergence semi-angle of 18.9 mrad and beam current of 15 pA. High-angle annular dark-field (HAADF) and low-angle ADF (LAADF) images are acquired with collection semi-angle ranges of 72-179 and 28-175 mrad, respectively. The dwell time is 2 ${\mu}$s. The effects of sample drift are corrected using the revolving STEM (revSTEM) method \cite{SANG201428} on a 12-frame 1024×1024 image series. For EDS, the electron probe current is 200 pA. A Thermo Fisher Scientific Super-X detector is used. 

Energy Dispersive X-ray Spectroscopy (EDS) quantification is carried out using calcium K lines, molybdenum L lines, and oxygen K lines, analyzed by the Thermo Fisher Scientific Velox software. A 1-sigma Gaussian postfilter is applied to reduce noise. 4D NBED datasets are collected using an Electron Microscope Pixel Array Detector (EMPAD) \cite{tate_high_2016}. The convergence semi-angle is 0.46 mrad. The representative NBED patterns for the film and substrate are averaged over corresponding regions with areas of approximately 1000$\times$150 nm$^2$ and 1000$\times$120 nm$^2$, respectively.

The optical characterization of the erbium emission using time-resolved photoluminescence excitation spectroscopy (PLE) and transient spectral holeburning (TSHB) is employed for the \ch{Er:CaMoO4} films and it is performed with samples mounted in a cryostat with a base temperature of 3.5 K. The details of the setup are described elsewhere~\cite{grant2024, ji2024}.

\section{Results and Discussion}

\subsection{Growth of stoichiometric \ch{Er:CaMoO4}}

 A preferred strategy for maintaining a 1:1 ratio for Ca and Mo in the growing oxide film is to identify and carry out the deposition in a regime where the growth is incorporation limited by one or both of the metallic species. In this regime, the metals incorporate only when the target compound, in this case \ch{CaMoO4} forms, but do not stick on the growing surface, or are not incorporated into other phases, otherwise. Hence, the first step is to determine the substrate temperatures beyond which the sticking coefficients of the calcium and molybdenum trioxide on the substrate surface are $\sim0$. 
 
 This is done by heating a Si substrate under \ch{Ca} or \ch{MoO3} molecular beams and monitoring the substrate surface RHEED pattern. For Ca, we observe that epitaxial Ca could be deposited below 550 \celsius{}, while above that temperature Ca does not stick to the substrate surface and a Si 2 x 1 surface reconstruction is retained. 
 
 For \ch{MoO3}, we observe polycrystalline growth below substrate temperatures of 400 \celsius{} (also confirmed by GIXRD). Above 400 \celsius{}, no film growth occurred, however the Si surface show signs of molybdenum silicide formation, as noted by changes in the RHEED pattern and the observation of the presence of Mo in surface from XPS measurements. Therefore, we conclude from these studies that a substrate temperature > 600 \celsius{} may be relevant for being in the incorporation limited regime for both materials. 

Our rationale in examining a growth space window for single phase \ch{CaMoO4} is as follows. With the \ch{Ca}, \ch{MoO3} and \ch{O2} beams on, there are three possible reactions at the surface of the substrate:

\begin{equation}
    2 \ch{MoO3} \rightleftharpoons 2 \ch{MoO2} + \ch{O2}
    \label{Eq1}
\end{equation}

\begin{equation}
    2 \ch{Ca} + \ch{O2} \rightarrow 2 \ch{CaO}
    \label{Eq2}
\end{equation}

\begin{equation}
    2 \ch{Ca} + 2 \ch{MoO3} + \ch{O2} \rightarrow 2 \ch{CaMoO4}
    \label{Eq3}
\end{equation}

The aim for incorporation limited growth of single phase \ch{CaMoO4} is to eliminate reactions~\ref{Eq1} and~\ref{Eq2} in the product direction, while promoting reaction~\ref{Eq3}. One way to accomplish this is by providing a high enough incoming flux of \ch{MoO3} and \ch{O2} (compared to the Ca flux), such that reaction~\ref{Eq3} is the dominant reaction with reaction rate controlled by the arrival rate of Ca adatoms on the growth surface. 

The substrate temperature is held high enough such that the only Mo incorporation into the film is enabled via reaction~\ref{Eq3}, and any free \ch{MoO3} adatoms on the growth surface desorbs (i.e. the sticking coefficient of \ch{MoO3} $\sim 0$). This prevents \ch{MoO3} formation on the growing surface. 

The possibility of formation of secondary phases of \ch{MoO2} (via reaction~\ref{Eq1}) complicates matters. However, since \ch{MoO2} is relatively non-volatile, this may be prevented by providing a combination of high enough \ch{O2} partial pressure and substrate temperature such that reaction~\ref{Eq1} is driven to the reactant side (see Supplemental material for equilibrium \ch{O2} partial pressures for reaction~\ref{Eq1}). We also need to ensure that reaction~\ref{Eq2} is suppressed to prevent non-volatile \ch{CaO} phase formation -- this can be accomplished by using a high \ch{MoO3}:\ch{Ca} ratio such that the reaction probability of reaction~\ref{Eq3} greatly exceeds that of reaction~\ref{Eq2}, i.e. conditions where a Ca adatom can always find adsorbed \ch{MoO3} and \ch{O2} on the surface within the Ca adatom migration length for reaction~\ref{Eq3} to proceed. 

Using this logical guidance we carried out a series of growth experiments at different \ch{MoO3}:\ch{Ca} ratios and \ch{O2} beam equivalent pressures with growth temperature of 650 \celsius{} on Si(100) substrates. 

The films were initially characterized using in-situ RHEED and GIXRD (following growth), to determine microstructure. The results show the presence of a single phase \ch{CaMoO4} growth window as indicated in Figure~\ref{SGW}(a) for growths performed at 650\celsius{}. It appears that such single phase growth requires a \ch{MoO3}/\ch{Ca} ratio of 2.9-3.5 and oxygen pressures of over 1.5e-5 torr (indicated by the green region in Figure~\ref{SGW}(a)). 

In this regime, only reaction~\ref{Eq3} persists because the oxygen pressure is high enough to prevent \ch{MoO2} formation described in equation~\ref{Eq1} and not enough spare calcium for reaction described by equation~\ref{Eq2} to occur. GIXRD spectra show a single phase thin film corresponding to \ch{CaMoO4} peaks (Figure~\ref{SGW}(b)), as identified in the figure. 

The multi-phase regimes are also identified in the figure. The regions shaded in black indicate growth parameters that lead to \ch{CaO} and \ch{CaMoO4} phases as identified by the GIXRD (Figure~\ref{SGW}(b)). This is due to the lack of molybdenum trioxide present at the surface of the substrate, leading to the formation of calcium oxide due to high oxygen pressure as described by equation~\ref{Eq2}. 

On the other hand, the area shaded in red yielded both \ch{MoO2} and \ch{CaMoO4} phases (corresponding GIXRD in Figure~\ref{SGW}(b)). This results from all calcium atoms reacting to \ch{MoO3} to form \ch{CaMoO4}, the rest of the molybdenum trioxide reduces to molybdenum dioxide due to an insufficient oxygen environment via chemical reaction described by equation~\ref{Eq1}. 

We do not observe growth window shifting between growths performed using molecular oxygen and atomic oxygen. 

While single phase growth of \ch{Er:CaMoO4} on Si substrates was achieved, the growths were polycrystalline due to the loss of epitaxial registry during the initial stages of the growth (as observed via in-situ RHEED studies during deposition). The likely cause of this is undesired interfaces forming on the substrate surface, leading to the loss of epitaxy. XPS analysis showed the formation of molybdenum disilicide and Magneli phases of molybdenum at the interface (see Supplemental Information), which are thought to disrupt further deposition of epitaxial material on the surface. 

\subsection{Microstructrual characterization of epitaxial \ch{Er:CaMoO4} on YSZ}

We now turn to epitaxial growth of these films on YSZ. We are able to grow single crystal epitaxial films of \ch{Er:CaMoO4} on YSZ substrates ($a=b=c=5.125$ \r{A}) in the incorporation limited regime identified in Figure~\ref{SGW}(a). Typical growth conditions are substrate temperature of 650-675$\degree$C, atomic oxygen supplied at $425W$ pressure of 2.3e-5 torr, and \ch{MoO3}/\ch{Ca} BEP ratio of $\sim3.4$ with total flux value ranging from 2e-7 to 1e-6 torr (50 nm/hr to 200 nm/hr). Growths are carried out with undoped buffer and capping layer to reduce interface effects on optical and magnetic properties ~\cite{singh2024}. 

Figure~\ref{EG}(a) shows a RHEED pattern (azimuth along YSZ <100>) of the \ch{CaMoO4} surface after the deposition of a 200 nm film. The streaky pattern indicates a smooth epitaxial surface. During the initial stages of the growth, the RHEED pattern smoothly transitions from a YSZ[001] to a \ch{CaMoO4}[001] pattern, suggesting good epitaxial alignment of YSZ[001]/\ch{CaMoO4}[001] and a layer by layer growth. This is further confirmed by X-ray diffraction studies. Figure~\ref{EG}(b) shows two $\omega-2\Theta$ HRXRD scans of as-grown (black curve) and annealed (green curve, annealed at 1000$\degree$C for 1 hour in \ch{Ar}/\ch{O2} (80/20\%) mix) \ch{CaMoO4} thin film on YSZ. The results are consistent with the epitaxial nature of the growth. The \ch{CaMoO4} (004) peak shows a linewidth of 375-380 arcseconds, with no significant improvement upon annealing but a slight shift of a $\sim$ 0.01\r{A}. The observed shifts reveals the change in lattice constant that is consistent with the relief of compressive in-plane strain. Relaxed lattice constant observed in this study (c = 11.418\r{A}) is $0.1\%$ smaller than the ones reported in bulk crystal and nanocrystals ~\cite{murai2023, hazen1985, rabuffetti2014, culver2014}. 

\begin{figure}
    \centering
    \includegraphics{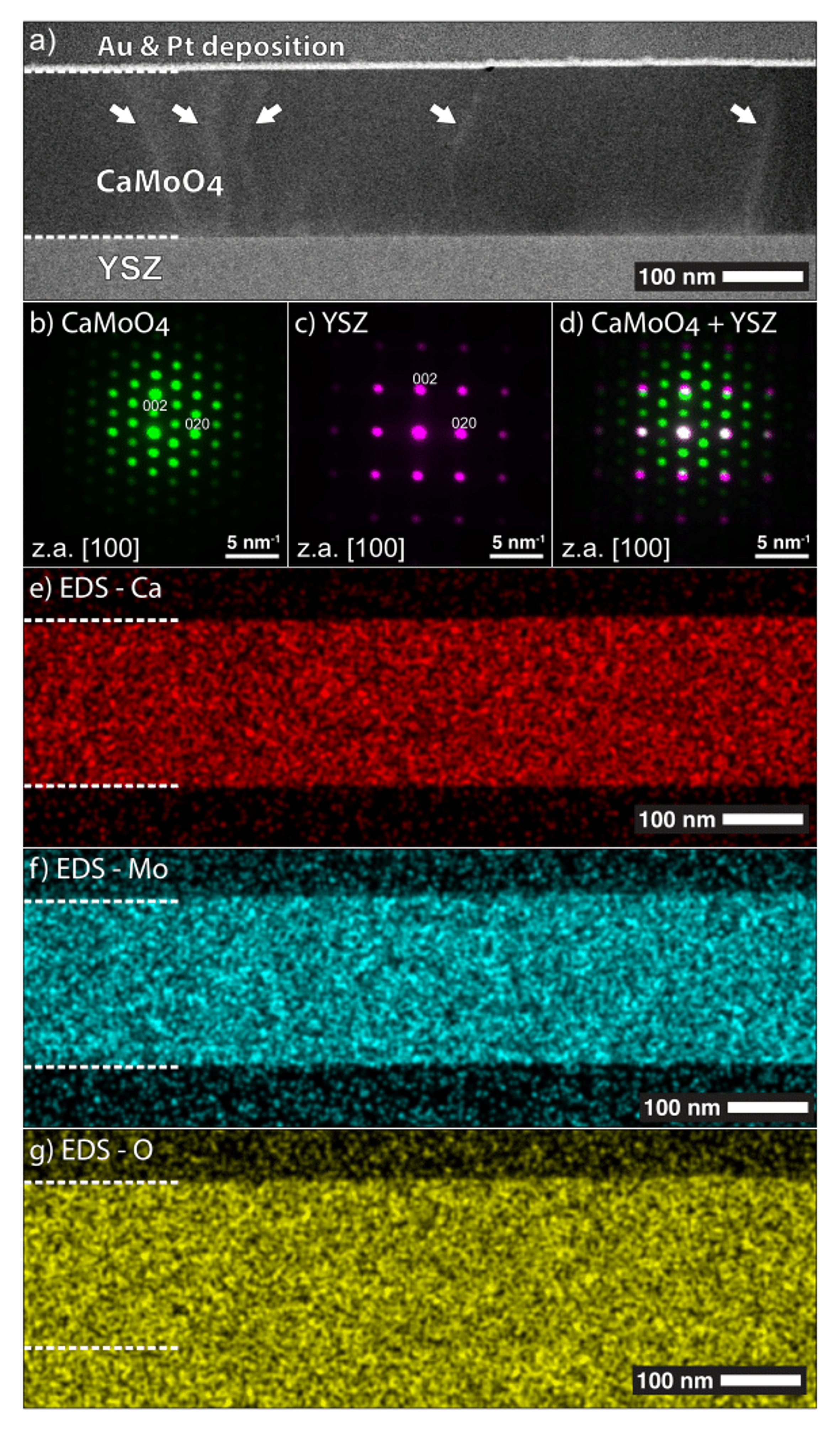}
    \caption{(a) LAADF-STEM image of an annealed (1000 \celsius{}, 1 hour, \ch{O2/Ar} 1 atm ambient) \ch{Er:CaMoO4} 200 nm 2 ppm film grown on YSZ(001) substrate. Threading dislocations (white arrows) are present at a $\sim10^9$ cm$^{-2}$ density throughout the film. (b-c) Diffraction patterns from the \ch{CaMoO4} film and YSZ substrate, averaged from the 4D NBED dataset along [100]. Their overlap (d) shows the epitaxial orientation relationship. (e-g) EDS mapping of calcium, molybdenum, and oxygen distribution from the same sampling region as (a). EDS mapping shows uniform distribution throughout the film and that a sharp interface between the film and the substrate is maintained after high-temperature annealing.}
    \label{EDS}
    
\end{figure}

\begin{figure}
    \centering
    \includegraphics{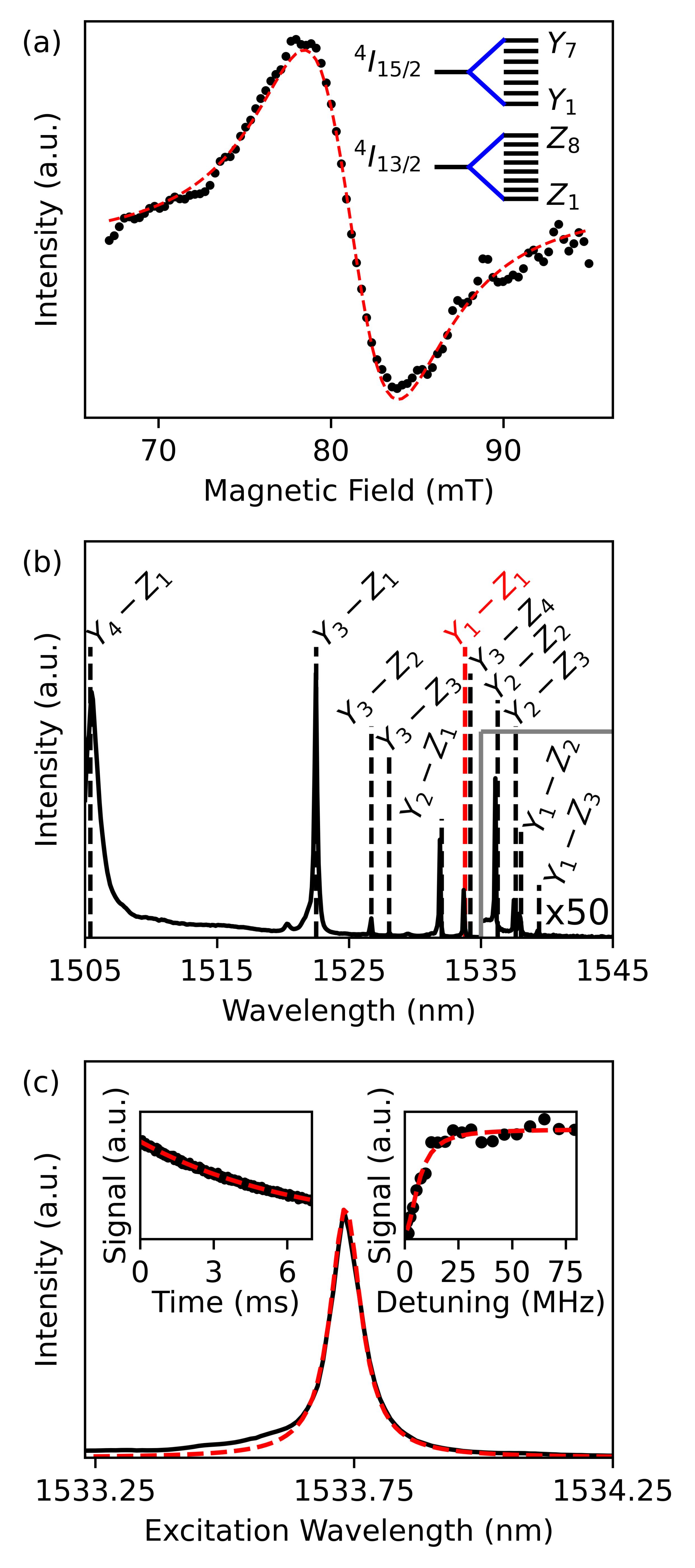}
    \caption{Characterization of 2 ppm annealed \ch{Er:CaMoO4} thin film on YSZ(001) substrate at 4 K. (a) CW EPR scan with a fit (red dashed line). The upper right inset shows the expected crystal field splitting for $^4I_{13/2} - ^4I_{15/2}$ transition of erbium incorporated in a tetragonal host. (b) A broad PLE scan of the sample performed at 4K. The transitions labeled with vertical dashed lines are found in literature~\cite{gerasivom2024}. The suspected $Y_1-Z_1$ transition is indicated with red dashed vertical line. (c) A narrow PLE scan of \ch{Er:CaMoO4} thin film, showing a suspected $Y_1-Z_1$ transition with a fit (red dashed line). The inset on the left show a time-resolved optical signature at the peak with a fit (red dashed line). The inset to the right is the transient spectral holeburning (TSHB) scan with a fit (red dashed line).}
    \label{Fig5}
 
\end{figure}

An atomic resolution HAADF-STEM image of the annealed \ch{CaMoO4}/\ch{YSZ} interface confirms the epitaxial registry (Figure~\ref{EG}(c)). Viewed along [100], the Ca and Mo atom columns are separated, with the Ca columns appearing dim and the Mo columns bright given the large atomic number difference. A schematic of the crystal structure is shown in the inset. Oxygen atoms are not resolved due to their low atomic number compared to the cations. The \ch{CaMoO4}/\ch{YSZ} interface is atomically sharp and flat, with steps within the range of one \ch{CaMoO4} unit cell. Misfit dislocations form at the interface to accommodate the lattice mismatch (+1.9\%) between the film and the substrate. 

Figure~\ref{EDS} provides a low magnification overview of the annealed \ch{Er:CaMoO4} film. The LAADF-STEM image (Figure~\ref{EDS}(a)) reveals threading dislocations in the epilayer with an approximate density of $10^9$ cm$^{-2}$. Diffraction patterns averaged from the 4D NBED dataset over the film and substrate (Figure~\ref{EDS}(b-d)) indicate that the film is single crystalline and epitaxial to the substrate with the following orientation relationship: $[100]_{\ch{YSZ}}\parallel[100]_{\ch{CaMoO4}}$ and $(001)_{\ch{YSZ}}\parallel(001)_{\ch{CaMoO4}}$, confirming RHEED observations. EDS maps acquired over the same region (Figure~\ref{EDS}(e-g)) shows that calcium, molybdenum, and oxygen are evenly distributed throughout the film, without phase separation or film/substrate intermixing. From EDS, the Ca to Mo atomic fraction ratio in the film is $1.03\pm0.15$, in agreement with the expected stoichiometric compound. 

\subsection{Optical and microwave studies of Er doped epitaxial \ch{CaMoO4}}

We now turn to the optical studies on the \ch{CaMoO4} epitaxial films doped with 2 ppm natural abundance Er. Films for optical studies were annealed at 1000\celsius{} for 1 hour in an \ch{Ar/O2} environment following MBE growth. It has been shown that erbium emission from an oxide thin film can be affected by performing such oxygen anneals~\cite{grant2024, clabel2015}. 

It is known that trivalent rare-earth ions incorporate into the Ca$^{2+}$ site for scheelites (\ch{ABO4}, where A = Ca and B = Mo, W)~\cite{ammerlan2001, kirton1965, mims1961}. Electron paramagnetic resonance (EPR) and optical studies of \ch{Er}$^{3+}$ in bulk \ch{CaMoO4} crystal have indicated that \ch{Er}$^{3+}$ ions are substitutionally incorporated into magnetically equivalent \ch{Ca}$^{2+}$ sites with $S_4$ local symmetry~\cite{gerasivom2024}. We would expect a similar substitutional incorporation of erbium into the \ch{CaMoO4} epitaxial films. Here, we carry out first EPR studies on the 2 ppm Er doped and annealed (1000\celsius{} for 1 hour in \ch{Ar/O2} mix) \ch{CaMoO4} thin film to confirm incorporation of Er in as-grown film. Figure~\ref{Fig5}(a) shows a CW EPR spectrum of the thin film sample at 4K, indicating a resonance occurring at about $81$ mT with a linewidth of $1.10(2)$ GHz. The observed resonance indicates a g-factor of $8.261(1)$, consistent with that observed for Er substitutionally incorporated into one of the atomic sites with S$_4$ symmetry in \ch{CaMoO4}~\cite{zverev1968, kurkin1970, gerasivom2024}. As such, the \ch{Er}$^{3+}$ ground state $^4I_{15/2}$ (referred to as Z) splits into 8 levels while the first excited state $^4I_{13/2}$ (referred to as Y) splits into 7 levels (as indicated in Figure~\ref{Fig5}(a) inset). The large spin inhomogeneous linewidths of EPR-allowed hyperfine transitions from \ch{Er}-167 ($I=7/2$) prevents us from prominently observing them. Both Er-Er, Er-grown-in defects interaction, interface and strain could lead to line broadening\cite{liu2005, grant2024, singh2024}. We can further reduce the spin inhomogeneous linewidth by lowering the doping of erbium as shown in other studies\cite{sati2007, grant2024}. 

Next, we turn to the on-resonant PLE measurements to probe the Er emission. Figure~\ref{Fig5}(b) shows a broad PLE spectrum ranging from 1505 to 1545 nm with 0.1 nm step size measured at 4K from the 2 ppm Er doped \ch{CaMoO4} on YSZ. We are able to observe multiple transitions. Comparing the observed transitions with the known $^4I_{13/2}$ to $^4I_{15/2}$ demonstrated from the bulk \ch{CaMoO4} (line labels)~\cite{gerasivom2024}, we find that there is good agreement with our measured result. We also observe two additional transitions, at 1521 nm and 1529 nm, not present in bulk \ch{Er:CaMoO4}. The physical origins of these are under further investigation. Both, PLE and EPR results are in agreement with values found in literature~\cite{gerasivom2024}, suggesting Er substituting Ca$^{2+}$ site with local $S_4$ symmetry. The comparison shown in Figure~\ref{Fig5}(b) also suggests that the $Y_1-Z_1$ optical transition is at $1533.73(1)$nm. Its location is in good agreement with reported values~\cite{gerasivom2024}. We further probe the $Y_1-Z_1$ transition at higher spectral resolution using PLE, as shown in Figure~\ref{Fig5}(c). The $Y_1-Z_1$ transition peak at $1533.73(1)$nm has an inhomogeneous linewidth of $9.1(1)$ GHz. The optical relaxation lifetime $T_1$ of this transition is $6.7(2)$ ms (left inset of Figure~\ref{Fig5}(c)). We also perform transient spectral holeburning (TSHB) measurement on this transition to probe spectral diffusion-limited homogeneous linewidth $\Gamma_{\text{SD}}$. We obtain $\Gamma_{SHB}$ of $6.7(4)$ MHz (Figure~\ref{Fig5}(c) right inset). 

Comparing these values to other $Y_1-Z_1$ transitions of erbium in various hosts with similar erbium concentration, $\Gamma_{inh}$ is comparable to the best reported in \ch{Er:CeO2} film~\cite{grant2024, zhang2024}, however, about a factor of 10 higher than those found in bulk \ch{TiO2} and \ch{CaWO4} crystals~\cite{ourari2023, phenicie2019}. That is likely due to the fact that bulk crystals are grown under near equilibrium conditions and hence likely have lower density of grown-in defects. $\Gamma_{SHB}$ is $\sim40\%$ higher than the best reported in \ch{Er:CeO2}, however, it is likely to be reduced by decreasing erbium doping~\cite{grant2024}. 

\section{Conclusion}

Erbium doped \ch{CaMoO4} crystal presents itself as a potential solid-state system with relatively low nuclear noise environment and with feasibility of epitaxial integration on silicon due to moderate lattice mismatch. In this study, we demonstrate the development of single phase \ch{CaMoO4} films on silicon and single phase epitaxial films on YSZ substrates. We also benchmark optical and microwave properties of single crystalline epitaxial \textit{in-situ} 2 ppm doped, 200 nm \ch{Er:CaMoO4} on YSZ: an optical inhomogeneous linewidth of 9.1(1) GHz, an optical relation time of 6.7(2) ms, a spectral diffusion-limited homogeneous linewidth of 6.7(4) MHz, and an EPR linewidth of 1.10(2) GHz. Future research is focused on reducing the erbium concentration in \ch{CaMoO4} films (as an approach for improving optical emission properties) as well as reducing the number of unintended defects, which would enable exploration of optical and spin coherence properties for quantum network applications.

\begin{acknowledgments}

This material is based upon work supported by the Air Force Office of Scientific Research under award number FA9550-24-1-0266. Part of this work was supported by advanced materials manufacturing technologies office and by Q-NEXT, a U.S. Department of Energy Office of Science National Quantum Information Science Research Centers under Award No. DE-FOA-0002253. Work performed at the Center for Nanoscale Materials, a U.S. Department of Energy Office of Science User Facility, was supported by the U.S. DOE, Office of Basic Energy Sciences, under Contract No. DE-AC02-06CH11357. SK, AYG, and JWE are grateful for funding from the DOE Advanced Materials and Manufacturing Technology Office (AMMTO) through the Energy Efficiency Scaling for Two Decades (EES2) program. Work at MIT was carried out using the facilities at MIT.nano and supported by the AFOSR through grant No. FA9550-23-1-0667.

\end{acknowledgments}

\section{Author Declarations}

\subsection*{Competing interests:}

All authors declare they have no competing interests.

\subsection*{Author Contributions}

I.M. and S.G. conceived of and designed the experiments. I.M. and W.L. carried out sample growths. B.Y.X.L., J.K., and J.M.L. carried out STEM and EDS characterization. I.M. carried out XRD measurements. S.K.S. and J.Z. carried out EPR measurements. G.D.G. and I.M. carried out SEM measurements. I.M. and G.D.G. carried out optical measurements with input from J.Z.. S.K., A.Y, and J. W. E. carried out XPS measurements. Overall interpretation and analysis of the results were led by I.M. and S.G. All authors contributed to the manuscript.

\textbf{Ignas Masiulionis}: Conceptualization (equal); Data curation (lead); Formal analysis (equal); Investigation (equal); Methodology (equal); Visualization (lead); Writing – original draft (equal); Writing – review \& editing (equal). \textbf{Bonnie Y.X. Lin}: Data curation (equal); Investigation (equal); Methodology (equal); Writing – original draft (equal); Writing – review \& editing (equal). \textbf{Sagar Kumar Seth}: Formal analysis (equal); Investigation (equal); Visualization (supporting). \textbf{Gregory D. Grant}: Formal analysis (equal); Methodology (equal); Visualization (equal). \textbf{Wanda Linquist}: Investigation (supporting). \textbf{Sungjoon Kim}: Data curation (supporting). \textbf{Junghwa Kim}: Investigation (supporting). \textbf{Angel Yanguas-Gil}: Investigation (supporting); \textbf{Jeffrey W. Elam}: Investigation (supporting); Resources (equal). \textbf{Jiefei Zhang}: Methodology (equal); Visualization (supporting); Writing – review \& editing (equal). \textbf{James M. LeBeau}: Investigation (equal); Formal analysis (equal); Funding acquisition (equal); Resources (equal); Supervision (equal); Writing – original draft (equal); Writing – review \& editing (equal). \textbf{David D. Awschalom}: Funding acquisition (equal); Methodology (equal); Resources (equal). \textbf{Supratik Guha}: Conceptualization (equal); Formal analysis (equal); Funding acquisition (equal); Methodology (equal); Resources (equal); Supervision (equal); Writing – original draft (equal); Writing – review \& editing (equal).

\section*{Data Availability Statement}

The data that support the findings of this study are available from the corresponding author upon reasonable request.

\nocite{*}


\begin{thebibliography}{30}%
	\makeatletter
	\providecommand \@ifxundefined [1]{%
		\@ifx{#1\undefined}
	}%
	\providecommand \@ifnum [1]{%
		\ifnum #1\expandafter \@firstoftwo
		\else \expandafter \@secondoftwo
		\fi
	}%
	\providecommand \@ifx [1]{%
		\ifx #1\expandafter \@firstoftwo
		\else \expandafter \@secondoftwo
		\fi
	}%
	\providecommand \natexlab [1]{#1}%
	\providecommand \enquote  [1]{``#1''}%
	\providecommand \bibnamefont  [1]{#1}%
	\providecommand \bibfnamefont [1]{#1}%
	\providecommand \citenamefont [1]{#1}%
	\providecommand \href@noop [0]{\@secondoftwo}%
	\providecommand \href [0]{\begingroup \@sanitize@url \@href}%
	\providecommand \@href[1]{\@@startlink{#1}\@@href}%
	\providecommand \@@href[1]{\endgroup#1\@@endlink}%
	\providecommand \@sanitize@url [0]{\catcode `\\12\catcode `\$12\catcode `\&12\catcode `\#12\catcode `\^12\catcode `\_12\catcode `\%12\relax}%
	\providecommand \@@startlink[1]{}%
	\providecommand \@@endlink[0]{}%
	\providecommand \url  [0]{\begingroup\@sanitize@url \@url }%
	\providecommand \@url [1]{\endgroup\@href {#1}{\urlprefix }}%
	\providecommand \urlprefix  [0]{URL }%
	\providecommand \Eprint [0]{\href }%
	\providecommand \doibase [0]{http://dx.doi.org/}%
	\providecommand \selectlanguage [0]{\@gobble}%
	\providecommand \bibinfo  [0]{\@secondoftwo}%
	\providecommand \bibfield  [0]{\@secondoftwo}%
	\providecommand \translation [1]{[#1]}%
	\providecommand \BibitemOpen [0]{}%
	\providecommand \bibitemStop [0]{}%
	\providecommand \bibitemNoStop [0]{.\EOS\space}%
	\providecommand \EOS [0]{\spacefactor3000\relax}%
	\providecommand \BibitemShut  [1]{\csname bibitem#1\endcsname}%
	\let\auto@bib@innerbib\@empty
	\bibitem [{\citenamefont {Wolfowicz}\ \emph {et~al.}(2021)\citenamefont {Wolfowicz}, \citenamefont {Heremans}, \citenamefont {Anderson}, \citenamefont {Kanai}, \citenamefont {Seo}, \citenamefont {Gali}, \citenamefont {Galli},\ and\ \citenamefont {Awschalom}}]{wolfowicz2021}%
	\BibitemOpen
	\bibfield  {author} {\bibinfo {author} {\bibfnamefont {G.}~\bibnamefont {Wolfowicz}}, \bibinfo {author} {\bibfnamefont {F.~J.}\ \bibnamefont {Heremans}}, \bibinfo {author} {\bibfnamefont {C.~P.}\ \bibnamefont {Anderson}}, \bibinfo {author} {\bibfnamefont {S.}~\bibnamefont {Kanai}}, \bibinfo {author} {\bibfnamefont {H.}~\bibnamefont {Seo}}, \bibinfo {author} {\bibfnamefont {A.}~\bibnamefont {Gali}}, \bibinfo {author} {\bibfnamefont {G.}~\bibnamefont {Galli}}, \ and\ \bibinfo {author} {\bibfnamefont {D.~D.}\ \bibnamefont {Awschalom}},\ }\bibfield  {title} {\enquote {\bibinfo {title} {Quantum guidelines for solid-state spin defects},}\ }\href@noop {} {\bibfield  {journal} {\bibinfo  {journal} {Nature Reviews Materials}\ }\textbf {\bibinfo {volume} {6}},\ \bibinfo {pages} {906--925} (\bibinfo {year} {2021})}\BibitemShut {NoStop}%
	\bibitem [{\citenamefont {Dibos}\ \emph {et~al.}(2022)\citenamefont {Dibos}, \citenamefont {Solomon}, \citenamefont {Sullivan}, \citenamefont {Singh}, \citenamefont {Sautter}, \citenamefont {Horn}, \citenamefont {Grant}, \citenamefont {Lin}, \citenamefont {Wen}, \citenamefont {Heremans} \emph {et~al.}}]{dibos2022}%
	\BibitemOpen
	\bibfield  {author} {\bibinfo {author} {\bibfnamefont {A.~M.}\ \bibnamefont {Dibos}}, \bibinfo {author} {\bibfnamefont {M.~T.}\ \bibnamefont {Solomon}}, \bibinfo {author} {\bibfnamefont {S.~E.}\ \bibnamefont {Sullivan}}, \bibinfo {author} {\bibfnamefont {M.~K.}\ \bibnamefont {Singh}}, \bibinfo {author} {\bibfnamefont {K.~E.}\ \bibnamefont {Sautter}}, \bibinfo {author} {\bibfnamefont {C.~P.}\ \bibnamefont {Horn}}, \bibinfo {author} {\bibfnamefont {G.~D.}\ \bibnamefont {Grant}}, \bibinfo {author} {\bibfnamefont {Y.}~\bibnamefont {Lin}}, \bibinfo {author} {\bibfnamefont {J.}~\bibnamefont {Wen}}, \bibinfo {author} {\bibfnamefont {F.~J.}\ \bibnamefont {Heremans}},  \emph {et~al.},\ }\bibfield  {title} {\enquote {\bibinfo {title} {Purcell enhancement of erbium ions in \ch{TiO2} on silicon nanocavities},}\ }\href@noop {} {\bibfield  {journal} {\bibinfo  {journal} {Nano Lett.}\ }\textbf {\bibinfo {volume} {22}},\ \bibinfo {pages} {6530--6536} (\bibinfo {year} {2022})}\BibitemShut {NoStop}%
	\bibitem [{\citenamefont {Ji}\ \emph {et~al.}(2024)\citenamefont {Ji}, \citenamefont {Solomon}, \citenamefont {Grant}, \citenamefont {Tanaka}, \citenamefont {Hua}, \citenamefont {Wen}, \citenamefont {Seth}, \citenamefont {Horn}, \citenamefont {Masiulionis}, \citenamefont {Singh} \emph {et~al.}}]{ji2024}%
	\BibitemOpen
	\bibfield  {author} {\bibinfo {author} {\bibfnamefont {C.}~\bibnamefont {Ji}}, \bibinfo {author} {\bibfnamefont {M.~T.}\ \bibnamefont {Solomon}}, \bibinfo {author} {\bibfnamefont {G.~D.}\ \bibnamefont {Grant}}, \bibinfo {author} {\bibfnamefont {K.}~\bibnamefont {Tanaka}}, \bibinfo {author} {\bibfnamefont {M.}~\bibnamefont {Hua}}, \bibinfo {author} {\bibfnamefont {J.}~\bibnamefont {Wen}}, \bibinfo {author} {\bibfnamefont {S.~K.}\ \bibnamefont {Seth}}, \bibinfo {author} {\bibfnamefont {C.~P.}\ \bibnamefont {Horn}}, \bibinfo {author} {\bibfnamefont {I.}~\bibnamefont {Masiulionis}}, \bibinfo {author} {\bibfnamefont {M.~K.}\ \bibnamefont {Singh}},  \emph {et~al.},\ }\bibfield  {title} {\enquote {\bibinfo {title} {Nanocavity-mediated purcell enhancement of \ch{Er} in \ch{TiO2} thin films grown via atomic layer deposition},}\ }\href@noop {} {\bibfield  {journal} {\bibinfo  {journal} {ACS Nano}\ }\textbf {\bibinfo {volume} {18}},\ \bibinfo {pages} {9929--9941} (\bibinfo {year} {2024})}\BibitemShut {NoStop}%
	\bibitem [{\citenamefont {Singh}\ \emph {et~al.}(2020)\citenamefont {Singh}, \citenamefont {Prakash}, \citenamefont {Wolfowicz}, \citenamefont {Wen}, \citenamefont {Huang}, \citenamefont {Rajh}, \citenamefont {Awschalom}, \citenamefont {Zhong},\ and\ \citenamefont {Guha}}]{singh2020}%
	\BibitemOpen
	\bibfield  {author} {\bibinfo {author} {\bibfnamefont {M.~K.}\ \bibnamefont {Singh}}, \bibinfo {author} {\bibfnamefont {A.}~\bibnamefont {Prakash}}, \bibinfo {author} {\bibfnamefont {G.}~\bibnamefont {Wolfowicz}}, \bibinfo {author} {\bibfnamefont {J.}~\bibnamefont {Wen}}, \bibinfo {author} {\bibfnamefont {Y.}~\bibnamefont {Huang}}, \bibinfo {author} {\bibfnamefont {T.}~\bibnamefont {Rajh}}, \bibinfo {author} {\bibfnamefont {D.~D.}\ \bibnamefont {Awschalom}}, \bibinfo {author} {\bibfnamefont {T.}~\bibnamefont {Zhong}}, \ and\ \bibinfo {author} {\bibfnamefont {S.}~\bibnamefont {Guha}},\ }\bibfield  {title} {\enquote {\bibinfo {title} {Epitaxial \ch{Er}-doped \ch{Y2O3} on silicon for quantum coherent devices},}\ }\href@noop {} {\bibfield  {journal} {\bibinfo  {journal} {APL Mat.}\ }\textbf {\bibinfo {volume} {8}} (\bibinfo {year} {2020})}\BibitemShut {NoStop}%
	\bibitem [{\citenamefont {Horvath}\ \emph {et~al.}(2023)\citenamefont {Horvath}, \citenamefont {Phenicie}, \citenamefont {Ourari}, \citenamefont {Uysal}, \citenamefont {Chen}, \citenamefont {Dusanowski}, \citenamefont {Raha}, \citenamefont {Stevenson}, \citenamefont {Turflinger}, \citenamefont {Cava} \emph {et~al.}}]{horvath2023}%
	\BibitemOpen
	\bibfield  {author} {\bibinfo {author} {\bibfnamefont {S.~P.}\ \bibnamefont {Horvath}}, \bibinfo {author} {\bibfnamefont {C.~M.}\ \bibnamefont {Phenicie}}, \bibinfo {author} {\bibfnamefont {S.}~\bibnamefont {Ourari}}, \bibinfo {author} {\bibfnamefont {M.~T.}\ \bibnamefont {Uysal}}, \bibinfo {author} {\bibfnamefont {S.}~\bibnamefont {Chen}}, \bibinfo {author} {\bibfnamefont {{\L}.}~\bibnamefont {Dusanowski}}, \bibinfo {author} {\bibfnamefont {M.}~\bibnamefont {Raha}}, \bibinfo {author} {\bibfnamefont {P.}~\bibnamefont {Stevenson}}, \bibinfo {author} {\bibfnamefont {A.~T.}\ \bibnamefont {Turflinger}}, \bibinfo {author} {\bibfnamefont {R.~J.}\ \bibnamefont {Cava}},  \emph {et~al.},\ }\bibfield  {title} {\enquote {\bibinfo {title} {Strong purcell enhancement of an optical magnetic dipole transition},}\ }\href@noop {} {\bibfield  {journal} {\bibinfo  {journal} {arXiv preprint arXiv:2307.03022}\ } (\bibinfo {year} {2023})}\BibitemShut {NoStop}%
	\bibitem [{\citenamefont {Rinner}\ \emph {et~al.}(2023)\citenamefont {Rinner}, \citenamefont {Burger}, \citenamefont {Gritsch}, \citenamefont {Schmitt},\ and\ \citenamefont {Reiserer}}]{rinner2023}%
	\BibitemOpen
	\bibfield  {author} {\bibinfo {author} {\bibfnamefont {S.}~\bibnamefont {Rinner}}, \bibinfo {author} {\bibfnamefont {F.}~\bibnamefont {Burger}}, \bibinfo {author} {\bibfnamefont {A.}~\bibnamefont {Gritsch}}, \bibinfo {author} {\bibfnamefont {J.}~\bibnamefont {Schmitt}}, \ and\ \bibinfo {author} {\bibfnamefont {A.}~\bibnamefont {Reiserer}},\ }\bibfield  {title} {\enquote {\bibinfo {title} {Erbium emitters in commercially fabricated nanophotonic silicon waveguides},}\ }\href@noop {} {\bibfield  {journal} {\bibinfo  {journal} {Nanophotonics}\ }\textbf {\bibinfo {volume} {12}},\ \bibinfo {pages} {3455--3462} (\bibinfo {year} {2023})}\BibitemShut {NoStop}%
	\bibitem [{\citenamefont {Zhang}\ \emph {et~al.}(2024)\citenamefont {Zhang}, \citenamefont {Grant}, \citenamefont {Masiulionis}, \citenamefont {Solomon}, \citenamefont {Marcks}, \citenamefont {Bindra}, \citenamefont {Niklas}, \citenamefont {Dibos}, \citenamefont {Poluektov}, \citenamefont {Heremans} \emph {et~al.}}]{zhang2024}%
	\BibitemOpen
	\bibfield  {author} {\bibinfo {author} {\bibfnamefont {J.}~\bibnamefont {Zhang}}, \bibinfo {author} {\bibfnamefont {G.~D.}\ \bibnamefont {Grant}}, \bibinfo {author} {\bibfnamefont {I.}~\bibnamefont {Masiulionis}}, \bibinfo {author} {\bibfnamefont {M.~T.}\ \bibnamefont {Solomon}}, \bibinfo {author} {\bibfnamefont {J.~C.}\ \bibnamefont {Marcks}}, \bibinfo {author} {\bibfnamefont {J.~K.}\ \bibnamefont {Bindra}}, \bibinfo {author} {\bibfnamefont {J.}~\bibnamefont {Niklas}}, \bibinfo {author} {\bibfnamefont {A.~M.}\ \bibnamefont {Dibos}}, \bibinfo {author} {\bibfnamefont {O.~G.}\ \bibnamefont {Poluektov}}, \bibinfo {author} {\bibfnamefont {F.~J.}\ \bibnamefont {Heremans}},  \emph {et~al.},\ }\bibfield  {title} {\enquote {\bibinfo {title} {Optical and spin coherence of \ch{Er} spin qubits in epitaxial cerium dioxide on silicon},}\ }\href@noop {} {\bibfield  {journal} {\bibinfo  {journal} {npj Quantum Inf.}\ }\textbf {\bibinfo {volume} {10}},\ \bibinfo {pages} {119} (\bibinfo {year} {2024})}\BibitemShut {NoStop}%
	\bibitem [{\citenamefont {Grant}\ \emph {et~al.}(2024)\citenamefont {Grant}, \citenamefont {Zhang}, \citenamefont {Masiulionis}, \citenamefont {Chattaraj}, \citenamefont {Sautter}, \citenamefont {Sullivan}, \citenamefont {Chebrolu}, \citenamefont {Liu}, \citenamefont {Martins}, \citenamefont {Niklas}, \citenamefont {Dibos}, \citenamefont {Kewalramani}, \citenamefont {Freeland}, \citenamefont {Wen}, \citenamefont {Poluektov}, \citenamefont {Heremans}, \citenamefont {Awschalom},\ and\ \citenamefont {Guha}}]{grant2024}%
	\BibitemOpen
	\bibfield  {author} {\bibinfo {author} {\bibfnamefont {G.~D.}\ \bibnamefont {Grant}}, \bibinfo {author} {\bibfnamefont {J.}~\bibnamefont {Zhang}}, \bibinfo {author} {\bibfnamefont {I.}~\bibnamefont {Masiulionis}}, \bibinfo {author} {\bibfnamefont {S.}~\bibnamefont {Chattaraj}}, \bibinfo {author} {\bibfnamefont {K.~E.}\ \bibnamefont {Sautter}}, \bibinfo {author} {\bibfnamefont {S.~E.}\ \bibnamefont {Sullivan}}, \bibinfo {author} {\bibfnamefont {R.}~\bibnamefont {Chebrolu}}, \bibinfo {author} {\bibfnamefont {Y.}~\bibnamefont {Liu}}, \bibinfo {author} {\bibfnamefont {J.~B.}\ \bibnamefont {Martins}}, \bibinfo {author} {\bibfnamefont {J.}~\bibnamefont {Niklas}}, \bibinfo {author} {\bibfnamefont {A.~M.}\ \bibnamefont {Dibos}}, \bibinfo {author} {\bibfnamefont {S.}~\bibnamefont {Kewalramani}}, \bibinfo {author} {\bibfnamefont {J.~W.}\ \bibnamefont {Freeland}}, \bibinfo {author} {\bibfnamefont {J.}~\bibnamefont {Wen}}, \bibinfo {author} {\bibfnamefont {O.~G.}\ \bibnamefont {Poluektov}}, \bibinfo {author}
		{\bibfnamefont {F.~J.}\ \bibnamefont {Heremans}}, \bibinfo {author} {\bibfnamefont {D.~D.}\ \bibnamefont {Awschalom}}, \ and\ \bibinfo {author} {\bibfnamefont {S.}~\bibnamefont {Guha}},\ }\bibfield  {title} {\enquote {\bibinfo {title} {Optical and microstructural characterization of $\ch{Er}^{3+}$ doped epitaxial cerium oxide on silicon},}\ }\href {\doibase 10.1063/5.0181717} {\bibfield  {journal} {\bibinfo  {journal} {APL Mater.}\ }\textbf {\bibinfo {volume} {12}},\ \bibinfo {pages} {021121} (\bibinfo {year} {2024})}\BibitemShut {NoStop}%
	\bibitem [{\citenamefont {Le~Dantec}\ \emph {et~al.}(2021)\citenamefont {Le~Dantec}, \citenamefont {Ran{\v{c}}i{\'c}}, \citenamefont {Lin}, \citenamefont {Billaud}, \citenamefont {Ranjan}, \citenamefont {Flanigan}, \citenamefont {Bertaina}, \citenamefont {Chaneli{\`e}re}, \citenamefont {Goldner}, \citenamefont {Erb} \emph {et~al.}}]{le2021}%
	\BibitemOpen
	\bibfield  {author} {\bibinfo {author} {\bibfnamefont {M.}~\bibnamefont {Le~Dantec}}, \bibinfo {author} {\bibfnamefont {M.}~\bibnamefont {Ran{\v{c}}i{\'c}}}, \bibinfo {author} {\bibfnamefont {S.}~\bibnamefont {Lin}}, \bibinfo {author} {\bibfnamefont {E.}~\bibnamefont {Billaud}}, \bibinfo {author} {\bibfnamefont {V.}~\bibnamefont {Ranjan}}, \bibinfo {author} {\bibfnamefont {D.}~\bibnamefont {Flanigan}}, \bibinfo {author} {\bibfnamefont {S.}~\bibnamefont {Bertaina}}, \bibinfo {author} {\bibfnamefont {T.}~\bibnamefont {Chaneli{\`e}re}}, \bibinfo {author} {\bibfnamefont {P.}~\bibnamefont {Goldner}}, \bibinfo {author} {\bibfnamefont {A.}~\bibnamefont {Erb}},  \emph {et~al.},\ }\bibfield  {title} {\enquote {\bibinfo {title} {Twenty-three--millisecond electron spin coherence of erbium ions in a natural-abundance crystal},}\ }\href@noop {} {\bibfield  {journal} {\bibinfo  {journal} {Sci. Adv.}\ }\textbf {\bibinfo {volume} {7}},\ \bibinfo {pages} {eabj9786} (\bibinfo {year} {2021})}\BibitemShut {NoStop}%
	\bibitem [{\citenamefont {Kanai}\ \emph {et~al.}(2022)\citenamefont {Kanai}, \citenamefont {Heremans}, \citenamefont {Seo}, \citenamefont {Wolfowicz}, \citenamefont {Anderson}, \citenamefont {Sullivan}, \citenamefont {Onizhuk}, \citenamefont {Galli}, \citenamefont {Awschalom},\ and\ \citenamefont {Ohno}}]{kanai2022}%
	\BibitemOpen
	\bibfield  {author} {\bibinfo {author} {\bibfnamefont {S.}~\bibnamefont {Kanai}}, \bibinfo {author} {\bibfnamefont {F.~J.}\ \bibnamefont {Heremans}}, \bibinfo {author} {\bibfnamefont {H.}~\bibnamefont {Seo}}, \bibinfo {author} {\bibfnamefont {G.}~\bibnamefont {Wolfowicz}}, \bibinfo {author} {\bibfnamefont {C.~P.}\ \bibnamefont {Anderson}}, \bibinfo {author} {\bibfnamefont {S.~E.}\ \bibnamefont {Sullivan}}, \bibinfo {author} {\bibfnamefont {M.}~\bibnamefont {Onizhuk}}, \bibinfo {author} {\bibfnamefont {G.}~\bibnamefont {Galli}}, \bibinfo {author} {\bibfnamefont {D.~D.}\ \bibnamefont {Awschalom}}, \ and\ \bibinfo {author} {\bibfnamefont {H.}~\bibnamefont {Ohno}},\ }\bibfield  {title} {\enquote {\bibinfo {title} {Generalized scaling of spin qubit coherence in over 12,000 host materials},}\ }\href {\doibase 10.1073/pnas.2121808119} {\bibfield  {journal} {\bibinfo  {journal} {Proc. Natl. Acad. Sci.}\ }\textbf {\bibinfo {volume} {119}},\ \bibinfo {pages} {e2121808119} (\bibinfo {year} {2022})}\BibitemShut
	{NoStop}%
	\bibitem [{\citenamefont {Ourari}\ \emph {et~al.}(2023)\citenamefont {Ourari}, \citenamefont {Dusanowski}, \citenamefont {Horvath}, \citenamefont {Uysal}, \citenamefont {Phenicie}, \citenamefont {Stevenson}, \citenamefont {Raha}, \citenamefont {Chen}, \citenamefont {Cava}, \citenamefont {de~Leon} \emph {et~al.}}]{ourari2023}%
	\BibitemOpen
	\bibfield  {author} {\bibinfo {author} {\bibfnamefont {S.}~\bibnamefont {Ourari}}, \bibinfo {author} {\bibfnamefont {{\L}.}~\bibnamefont {Dusanowski}}, \bibinfo {author} {\bibfnamefont {S.~P.}\ \bibnamefont {Horvath}}, \bibinfo {author} {\bibfnamefont {M.~T.}\ \bibnamefont {Uysal}}, \bibinfo {author} {\bibfnamefont {C.~M.}\ \bibnamefont {Phenicie}}, \bibinfo {author} {\bibfnamefont {P.}~\bibnamefont {Stevenson}}, \bibinfo {author} {\bibfnamefont {M.}~\bibnamefont {Raha}}, \bibinfo {author} {\bibfnamefont {S.}~\bibnamefont {Chen}}, \bibinfo {author} {\bibfnamefont {R.~J.}\ \bibnamefont {Cava}}, \bibinfo {author} {\bibfnamefont {N.~P.}\ \bibnamefont {de~Leon}},  \emph {et~al.},\ }\bibfield  {title} {\enquote {\bibinfo {title} {Indistinguishable telecom band photons from a single \ch{Er} ion in the solid state},}\ }\href@noop {} {\bibfield  {journal} {\bibinfo  {journal} {Nature}\ }\textbf {\bibinfo {volume} {620}},\ \bibinfo {pages} {977--981} (\bibinfo {year} {2023})}\BibitemShut {NoStop}%
	\bibitem [{\citenamefont {Gerasimov}\ \emph {et~al.}(2024)\citenamefont {Gerasimov}, \citenamefont {Baibekov}, \citenamefont {Minnegaliev}, \citenamefont {Shakurov}, \citenamefont {Zaripov}, \citenamefont {Moiseev}, \citenamefont {Lebedev},\ and\ \citenamefont {Malkin}}]{gerasivom2024}%
	\BibitemOpen
	\bibfield  {author} {\bibinfo {author} {\bibfnamefont {K.}~\bibnamefont {Gerasimov}}, \bibinfo {author} {\bibfnamefont {E.}~\bibnamefont {Baibekov}}, \bibinfo {author} {\bibfnamefont {M.}~\bibnamefont {Minnegaliev}}, \bibinfo {author} {\bibfnamefont {G.}~\bibnamefont {Shakurov}}, \bibinfo {author} {\bibfnamefont {R.}~\bibnamefont {Zaripov}}, \bibinfo {author} {\bibfnamefont {S.}~\bibnamefont {Moiseev}}, \bibinfo {author} {\bibfnamefont {A.}~\bibnamefont {Lebedev}}, \ and\ \bibinfo {author} {\bibfnamefont {B.}~\bibnamefont {Malkin}},\ }\bibfield  {title} {\enquote {\bibinfo {title} {Magneto-optical and high-frequency electron paramagnetic resonance spectroscopy of $\ch{Er}^{3+}$ ions in \ch{CaMoO4} single crystal},}\ }\href {\doibase https://doi.org/10.1016/j.jlumin.2024.120564} {\bibfield  {journal} {\bibinfo  {journal} {J. Lumin.}\ }\textbf {\bibinfo {volume} {270}},\ \bibinfo {pages} {120564} (\bibinfo {year} {2024})}\BibitemShut {NoStop}%
	\bibitem [{\citenamefont {ichiro Murai}\ \emph {et~al.}(2023)\citenamefont {ichiro Murai}, \citenamefont {Yamashita}, \citenamefont {Kitahara}, \citenamefont {Tokuda},\ and\ \citenamefont {Yoshiasa}}]{murai2023}%
	\BibitemOpen
	\bibfield  {author} {\bibinfo {author} {\bibfnamefont {K.}~\bibnamefont {ichiro Murai}}, \bibinfo {author} {\bibfnamefont {K.}~\bibnamefont {Yamashita}}, \bibinfo {author} {\bibfnamefont {G.}~\bibnamefont {Kitahara}}, \bibinfo {author} {\bibfnamefont {M.}~\bibnamefont {Tokuda}}, \ and\ \bibinfo {author} {\bibfnamefont {A.}~\bibnamefont {Yoshiasa}},\ }\bibfield  {title} {\enquote {\bibinfo {title} {Syntheses, single crystal structure analyses and ultraviolet light emission of $\ch{CaW_xMo_{(1-x)}O4}$ (x = 0.0-1.0) scheelite-powellite solid solutions},}\ }\href {\doibase 10.2465/jmps.220901} {\bibfield  {journal} {\bibinfo  {journal} {J. Mineral. Petrol. Sci.}\ }\textbf {\bibinfo {volume} {118}},\ \bibinfo {pages} {220901} (\bibinfo {year} {2023})}\BibitemShut {NoStop}%
	\bibitem [{\citenamefont {Reinhardt}\ and\ \citenamefont {Kern}(2018)}]{reinhardt2018}%
	\BibitemOpen
	\bibfield  {author} {\bibinfo {author} {\bibfnamefont {K.}~\bibnamefont {Reinhardt}}\ and\ \bibinfo {author} {\bibfnamefont {W.}~\bibnamefont {Kern}},\ }\href@noop {} {\emph {\bibinfo {title} {Handbook of silicon wafer cleaning technology}}}\ (\bibinfo  {publisher} {William Andrew},\ \bibinfo {year} {2018})\BibitemShut {NoStop}%
	\bibitem [{\citenamefont {Singh}\ \emph {et~al.}(2024)\citenamefont {Singh}, \citenamefont {Grant}, \citenamefont {Wolfowicz}, \citenamefont {Wen}, \citenamefont {Sullivan}, \citenamefont {Prakash}, \citenamefont {Dibos}, \citenamefont {Joseph~Heremans}, \citenamefont {Awschalom},\ and\ \citenamefont {Guha}}]{singh2024}%
	\BibitemOpen
	\bibfield  {author} {\bibinfo {author} {\bibfnamefont {M.~K.}\ \bibnamefont {Singh}}, \bibinfo {author} {\bibfnamefont {G.~D.}\ \bibnamefont {Grant}}, \bibinfo {author} {\bibfnamefont {G.}~\bibnamefont {Wolfowicz}}, \bibinfo {author} {\bibfnamefont {J.}~\bibnamefont {Wen}}, \bibinfo {author} {\bibfnamefont {S.~E.}\ \bibnamefont {Sullivan}}, \bibinfo {author} {\bibfnamefont {A.}~\bibnamefont {Prakash}}, \bibinfo {author} {\bibfnamefont {A.~M.}\ \bibnamefont {Dibos}}, \bibinfo {author} {\bibfnamefont {F.}~\bibnamefont {Joseph~Heremans}}, \bibinfo {author} {\bibfnamefont {D.~D.}\ \bibnamefont {Awschalom}}, \ and\ \bibinfo {author} {\bibfnamefont {S.}~\bibnamefont {Guha}},\ }\bibfield  {title} {\enquote {\bibinfo {title} {Optical and microstructural studies of erbium-doped \ch{TiO2} thin films on silicon, \ch{SrTiO3}, and sapphire},}\ }\href {\doibase 10.1063/5.0224010} {\bibfield  {journal} {\bibinfo  {journal} {J. Appl. Phys.}\ }\textbf {\bibinfo {volume} {136}},\ \bibinfo {pages} {124402} (\bibinfo {year}
		{2024})}\BibitemShut {NoStop}%
	\bibitem [{\citenamefont {Sang}\ and\ \citenamefont {LeBeau}(2014)}]{SANG201428}%
	\BibitemOpen
	\bibfield  {author} {\bibinfo {author} {\bibfnamefont {X.}~\bibnamefont {Sang}}\ and\ \bibinfo {author} {\bibfnamefont {J.~M.}\ \bibnamefont {LeBeau}},\ }\bibfield  {title} {\enquote {\bibinfo {title} {Revolving scanning transmission electron microscopy: Correcting sample drift distortion without prior knowledge},}\ }\href {\doibase https://doi.org/10.1016/j.ultramic.2013.12.004} {\bibfield  {journal} {\bibinfo  {journal} {Ultramicroscopy}\ }\textbf {\bibinfo {volume} {138}},\ \bibinfo {pages} {28--35} (\bibinfo {year} {2014})}\BibitemShut {NoStop}%
	\bibitem [{\citenamefont {Tate}\ \emph {et~al.}(2016)\citenamefont {Tate}, \citenamefont {Purohit}, \citenamefont {Chamberlain}, \citenamefont {Nguyen}, \citenamefont {Hovden}, \citenamefont {Chang}, \citenamefont {Deb}, \citenamefont {Turgut}, \citenamefont {Heron}, \citenamefont {Schlom}, \citenamefont {Ralph}, \citenamefont {Fuchs}, \citenamefont {Shanks}, \citenamefont {Philipp}, \citenamefont {Muller},\ and\ \citenamefont {Gruner}}]{tate_high_2016}%
	\BibitemOpen
	\bibfield  {author} {\bibinfo {author} {\bibfnamefont {M.~W.}\ \bibnamefont {Tate}}, \bibinfo {author} {\bibfnamefont {P.}~\bibnamefont {Purohit}}, \bibinfo {author} {\bibfnamefont {D.}~\bibnamefont {Chamberlain}}, \bibinfo {author} {\bibfnamefont {K.~X.}\ \bibnamefont {Nguyen}}, \bibinfo {author} {\bibfnamefont {R.}~\bibnamefont {Hovden}}, \bibinfo {author} {\bibfnamefont {C.~S.}\ \bibnamefont {Chang}}, \bibinfo {author} {\bibfnamefont {P.}~\bibnamefont {Deb}}, \bibinfo {author} {\bibfnamefont {E.}~\bibnamefont {Turgut}}, \bibinfo {author} {\bibfnamefont {J.~T.}\ \bibnamefont {Heron}}, \bibinfo {author} {\bibfnamefont {D.~G.}\ \bibnamefont {Schlom}}, \bibinfo {author} {\bibfnamefont {D.~C.}\ \bibnamefont {Ralph}}, \bibinfo {author} {\bibfnamefont {G.~D.}\ \bibnamefont {Fuchs}}, \bibinfo {author} {\bibfnamefont {K.~S.}\ \bibnamefont {Shanks}}, \bibinfo {author} {\bibfnamefont {H.~T.}\ \bibnamefont {Philipp}}, \bibinfo {author} {\bibfnamefont {D.~A.}\ \bibnamefont {Muller}}, \ and\ \bibinfo {author}
		{\bibfnamefont {S.~M.}\ \bibnamefont {Gruner}},\ }\bibfield  {title} {\enquote {\bibinfo {title} {High {Dynamic} {Range} {Pixel} {Array} {Detector} for {Scanning} {Transmission} {Electron} {Microscopy}},}\ }\href {\doibase 10.1017/S1431927615015664} {\bibfield  {journal} {\bibinfo  {journal} {Microsc. Microanal.}\ }\textbf {\bibinfo {volume} {22}},\ \bibinfo {pages} {237--249} (\bibinfo {year} {2016})}\BibitemShut {NoStop}%
	\bibitem [{\citenamefont {Hazen}, \citenamefont {Finger},\ and\ \citenamefont {Mariathasan}(1985)}]{hazen1985}%
	\BibitemOpen
	\bibfield  {author} {\bibinfo {author} {\bibfnamefont {R.~M.}\ \bibnamefont {Hazen}}, \bibinfo {author} {\bibfnamefont {L.~W.}\ \bibnamefont {Finger}}, \ and\ \bibinfo {author} {\bibfnamefont {J.~W.}\ \bibnamefont {Mariathasan}},\ }\bibfield  {title} {\enquote {\bibinfo {title} {High-pressure crystal chemistry of scheelite-type tungstates and molybdates},}\ }\href {\doibase https://doi.org/10.1016/0022-3697(85)90039-3} {\bibfield  {journal} {\bibinfo  {journal} {J. Phys. Chem. Solids}\ }\textbf {\bibinfo {volume} {46}},\ \bibinfo {pages} {253--263} (\bibinfo {year} {1985})}\BibitemShut {NoStop}%
	\bibitem [{\citenamefont {Rabuffetti}\ \emph {et~al.}(2014)\citenamefont {Rabuffetti}, \citenamefont {Culver}, \citenamefont {Suescun},\ and\ \citenamefont {Brutchey}}]{rabuffetti2014}%
	\BibitemOpen
	\bibfield  {author} {\bibinfo {author} {\bibfnamefont {F.~A.}\ \bibnamefont {Rabuffetti}}, \bibinfo {author} {\bibfnamefont {S.~P.}\ \bibnamefont {Culver}}, \bibinfo {author} {\bibfnamefont {L.}~\bibnamefont {Suescun}}, \ and\ \bibinfo {author} {\bibfnamefont {R.~L.}\ \bibnamefont {Brutchey}},\ }\bibfield  {title} {\enquote {\bibinfo {title} {Structural disorder in \ch{AMoO4} (\ch{A} = \ch{Ca}, \ch{Sr}, \ch{Ba}) scheelite nanocrystals},}\ }\href {\doibase 10.1021/ic4025348} {\bibfield  {journal} {\bibinfo  {journal} {Inorg. Chem.}\ }\textbf {\bibinfo {volume} {53}},\ \bibinfo {pages} {1056--1061} (\bibinfo {year} {2014})},\ \bibinfo {note} {pMID: 24266706}\BibitemShut {NoStop}%
	\bibitem [{\citenamefont {Culver}\ \emph {et~al.}(2013)\citenamefont {Culver}, \citenamefont {Rabuffetti}, \citenamefont {Zhou}, \citenamefont {Mecklenburg}, \citenamefont {Song}, \citenamefont {Melot},\ and\ \citenamefont {Brutchey}}]{culver2014}%
	\BibitemOpen
	\bibfield  {author} {\bibinfo {author} {\bibfnamefont {S.~P.}\ \bibnamefont {Culver}}, \bibinfo {author} {\bibfnamefont {F.~A.}\ \bibnamefont {Rabuffetti}}, \bibinfo {author} {\bibfnamefont {S.}~\bibnamefont {Zhou}}, \bibinfo {author} {\bibfnamefont {M.}~\bibnamefont {Mecklenburg}}, \bibinfo {author} {\bibfnamefont {Y.}~\bibnamefont {Song}}, \bibinfo {author} {\bibfnamefont {B.~C.}\ \bibnamefont {Melot}}, \ and\ \bibinfo {author} {\bibfnamefont {R.~L.}\ \bibnamefont {Brutchey}},\ }\bibfield  {title} {\enquote {\bibinfo {title} {Low-temperature synthesis of \ch{AMoO4} (\ch{A} = \ch{Ca}, \ch{Sr}, \ch{Ba}) scheelite nanocrystals},}\ }\href {\doibase 10.1021/cm402867y} {\bibfield  {journal} {\bibinfo  {journal} {Chem. Mater.}\ }\textbf {\bibinfo {volume} {25}},\ \bibinfo {pages} {4129--4134} (\bibinfo {year} {2013})}\BibitemShut {NoStop}%
	\bibitem [{\citenamefont {{Clabel H.}}\ \emph {et~al.}(2015)\citenamefont {{Clabel H.}}, \citenamefont {Rivera}, \citenamefont {{Siu Li}}, \citenamefont {Nunes}, \citenamefont {Leite}, \citenamefont {Schreiner},\ and\ \citenamefont {Marega}}]{clabel2015}%
	\BibitemOpen
	\bibfield  {author} {\bibinfo {author} {\bibfnamefont {J.}~\bibnamefont {{Clabel H.}}}, \bibinfo {author} {\bibfnamefont {V.}~\bibnamefont {Rivera}}, \bibinfo {author} {\bibfnamefont {M.}~\bibnamefont {{Siu Li}}}, \bibinfo {author} {\bibfnamefont {L.}~\bibnamefont {Nunes}}, \bibinfo {author} {\bibfnamefont {E.}~\bibnamefont {Leite}}, \bibinfo {author} {\bibfnamefont {W.}~\bibnamefont {Schreiner}}, \ and\ \bibinfo {author} {\bibfnamefont {E.}~\bibnamefont {Marega}},\ }\bibfield  {title} {\enquote {\bibinfo {title} {Near-infrared light emission of \ch{Er}$^{3+}$-doped zirconium oxide thin films: An optical, structural and {XPS} study},}\ }\href {\doibase https://doi.org/10.1016/j.jallcom.2014.09.007} {\bibfield  {journal} {\bibinfo  {journal} {J. Alloys Compd.}\ }\textbf {\bibinfo {volume} {619}},\ \bibinfo {pages} {800--806} (\bibinfo {year} {2015})}\BibitemShut {NoStop}%
	\bibitem [{\citenamefont {Ammerlaan}\ and\ \citenamefont {de~Maat-Gersdorf}(2021)}]{ammerlan2001}%
	\BibitemOpen
	\bibfield  {author} {\bibinfo {author} {\bibfnamefont {C.}~\bibnamefont {Ammerlaan}}\ and\ \bibinfo {author} {\bibfnamefont {I.}~\bibnamefont {de~Maat-Gersdorf}},\ }\bibfield  {title} {\enquote {\bibinfo {title} {Zeeman splitting factor of the \ch{Er}$^{3+}$ ion in a crystal field},}\ }\href {\doibase https://doi.org/10.1007/BF03162436} {\bibfield  {journal} {\bibinfo  {journal} {Appl. Magn. Reson.}\ }\textbf {\bibinfo {volume} {21}},\ \bibinfo {pages} {13--33} (\bibinfo {year} {2021})}\BibitemShut {NoStop}%
	\bibitem [{\citenamefont {Kirton}(1965)}]{kirton1965}%
	\BibitemOpen
	\bibfield  {author} {\bibinfo {author} {\bibfnamefont {J.}~\bibnamefont {Kirton}},\ }\bibfield  {title} {\enquote {\bibinfo {title} {Paramagnetic resonance of trivalent holmium ions in calcium tungstate},}\ }\href {\doibase 10.1103/PhysRev.139.A1930} {\bibfield  {journal} {\bibinfo  {journal} {Phys. Rev.}\ }\textbf {\bibinfo {volume} {139}},\ \bibinfo {pages} {A1930--A1933} (\bibinfo {year} {1965})}\BibitemShut {NoStop}%
	\bibitem [{\citenamefont {Mims}, \citenamefont {Nassau},\ and\ \citenamefont {McGee}(1961)}]{mims1961}%
	\BibitemOpen
	\bibfield  {author} {\bibinfo {author} {\bibfnamefont {W.~B.}\ \bibnamefont {Mims}}, \bibinfo {author} {\bibfnamefont {K.}~\bibnamefont {Nassau}}, \ and\ \bibinfo {author} {\bibfnamefont {J.~D.}\ \bibnamefont {McGee}},\ }\bibfield  {title} {\enquote {\bibinfo {title} {Spectral diffusion in electron resonance lines},}\ }\href {\doibase 10.1103/PhysRev.123.2059} {\bibfield  {journal} {\bibinfo  {journal} {Phys. Rev.}\ }\textbf {\bibinfo {volume} {123}},\ \bibinfo {pages} {2059--2069} (\bibinfo {year} {1961})}\BibitemShut {NoStop}%
	\bibitem [{\citenamefont {G.M.~Zverev}(1968)}]{zverev1968}%
	\BibitemOpen
	\bibfield  {author} {\bibinfo {author} {\bibfnamefont {A.~S.}\ \bibnamefont {G.M.~Zverev}, \bibfnamefont {L.V.~Makarenko}},\ }\bibfield  {title} {\enquote {\bibinfo {title} {{EPR} of ground and excited states of the \ch{Er}$^{3+}$ ion in crystals of calcium and strontium molybdates},}\ }\href@noop {} {\bibfield  {journal} {\bibinfo  {journal} {Sov. Phys. Solid State}\ }\textbf {\bibinfo {volume} {9}},\ \bibinfo {pages} {3651--3653} (\bibinfo {year} {1968})}\BibitemShut {NoStop}%
	\bibitem [{\citenamefont {I.N.~Kurkin}(1970)}]{kurkin1970}%
	\BibitemOpen
	\bibfield  {author} {\bibinfo {author} {\bibfnamefont {E.~T.}\ \bibnamefont {I.N.~Kurkin}},\ }\bibfield  {title} {\enquote {\bibinfo {title} {Spin-lattice relaxation of \ch{Er}$^{3+}$ ions in crystals of the homologous scheelite series},}\ }\href@noop {} {\bibfield  {journal} {\bibinfo  {journal} {Sov. Phys. Solid State}\ }\textbf {\bibinfo {volume} {11}},\ \bibinfo {pages} {3027--3029} (\bibinfo {year} {1970})}\BibitemShut {NoStop}%
	\bibitem [{\citenamefont {{Liu}}\ and\ \citenamefont {{Jacquier}}(2005)}]{liu2005}%
	\BibitemOpen
	\bibfield  {author} {\bibinfo {author} {\bibfnamefont {G.}~\bibnamefont {{Liu}}}\ and\ \bibinfo {author} {\bibfnamefont {B.}~\bibnamefont {{Jacquier}}},\ }\href@noop {} {\emph {\bibinfo {title} {{Spectroscopic Properties of Rare Earths in Optical Materials}}}}\ (\bibinfo  {publisher} {Springer},\ \bibinfo {year} {2005})\BibitemShut {NoStop}%
	\bibitem [{\citenamefont {Sati}, \citenamefont {Stepanov},\ and\ \citenamefont {Pashchenko}(2007)}]{sati2007}%
	\BibitemOpen
	\bibfield  {author} {\bibinfo {author} {\bibfnamefont {P.}~\bibnamefont {Sati}}, \bibinfo {author} {\bibfnamefont {A.}~\bibnamefont {Stepanov}}, \ and\ \bibinfo {author} {\bibfnamefont {V.}~\bibnamefont {Pashchenko}},\ }\bibfield  {title} {\enquote {\bibinfo {title} {Exchange broadening of {EPR} line in \ch{ZnO:Co}},}\ }\href {\doibase 10.1063/1.2747067} {\bibfield  {journal} {\bibinfo  {journal} {Low Temp. Phys.}\ }\textbf {\bibinfo {volume} {33}},\ \bibinfo {pages} {927--930} (\bibinfo {year} {2007})}\BibitemShut {NoStop}%
	\bibitem [{\citenamefont {Phenicie}\ \emph {et~al.}(2019)\citenamefont {Phenicie}, \citenamefont {Stevenson}, \citenamefont {Welinski}, \citenamefont {Rose}, \citenamefont {Asfaw}, \citenamefont {Cava}, \citenamefont {Lyon}, \citenamefont {de~Leon},\ and\ \citenamefont {Thompson}}]{phenicie2019}%
	\BibitemOpen
	\bibfield  {author} {\bibinfo {author} {\bibfnamefont {C.~M.}\ \bibnamefont {Phenicie}}, \bibinfo {author} {\bibfnamefont {P.}~\bibnamefont {Stevenson}}, \bibinfo {author} {\bibfnamefont {S.}~\bibnamefont {Welinski}}, \bibinfo {author} {\bibfnamefont {B.~C.}\ \bibnamefont {Rose}}, \bibinfo {author} {\bibfnamefont {A.~T.}\ \bibnamefont {Asfaw}}, \bibinfo {author} {\bibfnamefont {R.~J.}\ \bibnamefont {Cava}}, \bibinfo {author} {\bibfnamefont {S.~A.}\ \bibnamefont {Lyon}}, \bibinfo {author} {\bibfnamefont {N.~P.}\ \bibnamefont {de~Leon}}, \ and\ \bibinfo {author} {\bibfnamefont {J.~D.}\ \bibnamefont {Thompson}},\ }\bibfield  {title} {\enquote {\bibinfo {title} {Narrow optical line widths in erbium implanted in \ch{TiO2}},}\ }\href {\doibase 10.1021/acs.nanolett.9b03831} {\bibfield  {journal} {\bibinfo  {journal} {Nano Letters}\ }\textbf {\bibinfo {volume} {19}},\ \bibinfo {pages} {8928--8933} (\bibinfo {year} {2019})}\BibitemShut {NoStop}%
	\bibitem [{\citenamefont {Parchur}\ and\ \citenamefont {Ningthoujam}(2011)}]{parchur2011}%
	\BibitemOpen
	\bibfield  {author} {\bibinfo {author} {\bibfnamefont {A.~K.}\ \bibnamefont {Parchur}}\ and\ \bibinfo {author} {\bibfnamefont {R.~S.}\ \bibnamefont {Ningthoujam}},\ }\bibfield  {title} {\enquote {\bibinfo {title} {Preparation and structure refinement of \ch{Eu}$^{3+}$ doped \ch{CaMoO4} nanoparticles},}\ }\href@noop {} {\bibfield  {journal} {\bibinfo  {journal} {Dalton Trans.}\ }\textbf {\bibinfo {volume} {40}},\ \bibinfo {pages} {7590--7594} (\bibinfo {year} {2011})}\BibitemShut {NoStop}%
\end{thebibliography}
\end{document}